\documentclass[12pt]{article}

\usepackage{graphicx}
\usepackage{amsmath}
\usepackage{amsfonts}
\usepackage{amssymb}
\usepackage{hyperref}
\usepackage{cite}
\usepackage{float}
\usepackage{multirow}

\title{A Novel Approach to Image Steganography Using Generative Adversarial Networks}
\author{Waheed Rehman}
\date{\today}

\begin{document}

\maketitle

\begin{abstract}
The field of steganography has long been focused on developing methods to securely embed information within various digital media while ensuring imperceptibility and robustness. However, the growing sophistication of detection tools and the demand for increased data hiding capacity have revealed limitations in traditional techniques. In this paper, we propose a novel approach to image steganography that leverages the power of generative adversarial networks (GANs) to address these challenges. By employing a carefully designed GAN architecture, our method ensures the creation of stego-images that are visually indistinguishable from their original counterparts, effectively thwarting detection by advanced steganalysis tools. Additionally, the adversarial training paradigm optimizes the balance between embedding capacity, imperceptibility, and robustness, enabling more efficient and secure data hiding. We evaluate our proposed method through a series of experiments on benchmark datasets and compare its performance against baseline techniques, including least significant bit (LSB) substitution and discrete cosine transform (DCT)-based methods. Our results demonstrate significant improvements in metrics such as Peak Signal-to-Noise Ratio (PSNR), Structural Similarity Index Measure (SSIM), and robustness against detection. This work not only contributes to the advancement of image steganography but also provides a foundation for exploring GAN-based approaches for secure digital communication.
\end{abstract}

\section{Introduction}

Deep learning has revolutionized the field of computer vision, enabling unprecedented advancements in tasks such as image classification~\cite{colornet,resnet}, action recognition~\cite{gowda2017human,simonyan2014two}, and generative modeling~\cite{gans1}. With the advent of convolutional neural networks (CNNs)~\cite{resnet} and generative adversarial networks (GANs)~\cite{gans1}, deep learning has provided powerful tools to process and generate visual data with remarkable accuracy and realism. These breakthroughs have not only pushed the boundaries of traditional computer vision applications but also opened new possibilities in niche areas such as image synthesis, style transfer, and secure data embedding, including steganography.

Steganography, the practice of concealing information within digital media, has been an area of active research for decades. Derived from the Greek words "steganos" (covered) and "graphy" (writing), steganography focuses on enabling covert communication by embedding secret data within a medium, such as images, audio, or video. Among these, image steganography has gained significant attention due to the prevalence and versatility of digital images in modern communication systems \cite{johnson2001exploring,gowda1}.

The primary goals of image steganography are to achieve high imperceptibility, robustness, and embedding capacity. Imperceptibility ensures that the modifications made to the cover image are not noticeable to human vision or statistical analysis. Robustness guarantees that the embedded data remains intact and retrievable even after undergoing common image processing operations such as compression, scaling, or noise addition. Embedding capacity refers to the amount of data that can be securely hidden without compromising imperceptibility or robustness \cite{provos2003hide,gowda2}.

\subsection{Challenges in Image Steganography}
Traditional steganography methods often struggle to balance the trade-offs among imperceptibility, robustness, and embedding capacity. Spatial domain techniques, such as least significant bit (LSB) substitution, are computationally efficient and simple but are vulnerable to detection and distortion under image manipulations \cite{chan2004hiding}. Transform domain methods, which embed data in the frequency components of an image, offer greater robustness but require higher computational resources and exhibit limited embedding capacity \cite{chen2006highly,gowda3}.

Furthermore, the increasing sophistication of steganalysis tools poses significant challenges to traditional methods. Steganalysis, the science of detecting hidden data, has leveraged machine learning and deep learning techniques to identify subtle patterns introduced by embedding schemes \cite{qian2015deep,gowda6}. As a result, developing steganographic systems that can evade detection while maintaining robustness has become more complex.

\subsection{Motivation for Using Deep Learning}
Deep learning, particularly convolutional neural networks (CNNs) and generative adversarial networks (GANs), has revolutionized numerous fields, including computer vision, natural language processing, and cybersecurity. These advancements have also influenced the domain of steganography, where deep learning models are increasingly being used to optimize data embedding and extraction processes.

CNNs have been employed for both steganographic embedding and steganalysis. For instance, Baluja \cite{baluja2017hiding} introduced a deep learning framework for hiding one image within another, demonstrating improved robustness and imperceptibility. Meanwhile, adversarial approaches using GANs have shown significant promise by enabling the generation of stego-images that are indistinguishable from cover images \cite{zhang2019steganogan}. GANs use a generator-discriminator framework where the generator learns to embed data while the discriminator attempts to distinguish between cover and stego-images, driving the system to produce more realistic and undetectable stego-images.

\subsection{Contributions of This Work}
This paper proposes a novel approach to image steganography using generative adversarial networks (GANs). The primary contributions of this work are as follows:
\begin{itemize}
    \item We design a GAN-based framework that optimizes imperceptibility, robustness, and embedding capacity simultaneously, addressing the limitations of traditional and existing deep learning-based methods.
    \item We introduce a loss function tailored for steganography that balances adversarial training with reconstruction accuracy, ensuring both high-quality stego-images and reliable data extraction.
    \item We evaluate the proposed method against baseline techniques, including least significant bit (LSB) substitution and discrete cosine transform (DCT)-based embedding, using standard metrics such as PSNR, SSIM, and detection accuracy.
\end{itemize}

The results demonstrate that our approach outperforms existing methods, offering improved security and efficiency for covert communication. This work not only advances the state of the art in image steganography but also highlights the potential of GANs for secure digital communication.

\section{Related Work}
\label{sec:related_work}
Image steganography has been extensively studied, with research spanning traditional techniques, machine learning-based methods, and the emerging application of generative adversarial networks (GANs). This section provides a review of these approaches, focusing on their contributions and limitations.

\subsection{Traditional Image Steganography Techniques}
Traditional methods for image steganography are generally categorized into spatial domain techniques and transform domain techniques.

\subsubsection{Spatial Domain Techniques}
Spatial domain techniques directly modify the pixel values of the cover image to embed secret data. The most notable approach in this category is the least significant bit (LSB) substitution, which involves altering the least significant bits of pixel values to encode data \cite{chan2004hiding,lsb1,gowda7,lsb2}. While LSB substitution is computationally efficient and easy to implement, it is highly vulnerable to statistical attacks and visual inspection.

Other advancements in spatial domain steganography include edge-based embedding techniques, which focus on embedding data in high-gradient regions of the image to reduce perceptual distortion \cite{moulin2003information,ghosal2021image,gowda4}. However, these methods often suffer from limited embedding capacity and poor robustness to image processing operations.

\subsubsection{Transform Domain Techniques}
Transform domain techniques embed secret data into the frequency components of the image, offering better robustness to compression and noise. Discrete cosine transform (DCT) and discrete wavelet transform (DWT) are the most commonly used transform domain methods. In DCT-based techniques, data is embedded in the middle-frequency coefficients, balancing imperceptibility and robustness \cite{chen2006highly,gowda5,raja2005secure}. DWT-based methods further enhance robustness by leveraging the multi-resolution properties of wavelets \cite{sinha2009digital,po2006dwt}.

Although transform domain techniques are more robust than spatial domain methods, they often require higher computational resources and exhibit trade-offs between embedding capacity and imperceptibility.

\subsection{Machine Learning-Based Steganography}
With the advent of deep learning, machine learning-based methods have emerged as a powerful alternative to traditional approaches. These methods employ data-driven models to optimize the embedding and extraction processes.

\subsubsection{Convolutional Neural Networks (CNNs) for Steganography}
Several studies have explored the use of convolutional neural networks (CNNs) for image steganography. For instance, Baluja \cite{baluja2017hiding} proposed a deep learning framework where a CNN is used for both data embedding and extraction. The method demonstrated improved imperceptibility and robustness compared to traditional techniques. However, the approach required significant computational resources for training and inference.

\subsubsection{Adversarial Attacks and Steganalysis}
In parallel, CNNs have been employed for steganalysis, the process of detecting hidden data within images. These advancements have posed significant challenges to traditional steganography methods, necessitating the development of more robust techniques \cite{qian2015deep}.

\subsection{Generative Adversarial Networks (GANs) in Steganography}
Generative adversarial networks (GANs) have recently been adopted for image steganography, offering a novel approach to optimize imperceptibility and robustness. GANs consist of two networks: a generator, which embeds the secret data, and a discriminator, which aims to distinguish between cover and stego-images.

\subsubsection{GAN-Based Methods}
Volkhonskiy et al. \cite{volkhonskiy2020steganographic} introduced the use of GANs to generate stego-images that are visually indistinguishable from cover images. Their approach demonstrated significant improvements in imperceptibility but faced challenges in maintaining high embedding capacity.

Zhang et al. \cite{zhang2019steganogan} proposed SteganoGAN, a GAN-based framework that combines adversarial training with loss functions tailored for steganography. SteganoGAN achieved state-of-the-art performance in imperceptibility and robustness but required careful tuning of the model parameters.

\subsubsection{Challenges and Limitations}
Despite their advantages, GAN-based methods face challenges such as high computational complexity and instability during training. Additionally, their performance is often dataset-dependent, limiting their generalizability.

\subsection{Summary of Related Work}
In summary, traditional methods provide a strong foundation for steganography but face limitations in robustness and embedding capacity. Machine learning-based approaches, particularly GANs, offer promising solutions to these challenges. However, further research is needed to address issues such as computational efficiency and generalizability.

\section{Proposed Method}
\label{sec:proposed_method}
This section introduces the proposed generative adversarial network (GAN)-based framework for image steganography. The framework is designed to achieve a superior balance among imperceptibility, robustness, and embedding capacity, addressing the limitations of traditional methods and previous GAN-based approaches.

\subsection{Framework Overview}
The proposed method utilizes a GAN architecture comprising three components: the \textit{generator}, the \textit{discriminator}, and the \textit{extractor}. These components work collaboratively to embed secret data into a cover image, ensuring that the resulting stego-image is visually indistinguishable from the cover image while enabling accurate retrieval of the hidden data.

\begin{itemize}
    \item \textbf{Generator ($G$):} Embeds the secret data into the cover image to produce the stego-image.
    \item \textbf{Discriminator ($D$):} Differentiates between cover and stego-images, guiding the generator to produce high-quality outputs.
    \item \textbf{Extractor ($E$):} Recovers the secret data from the stego-image, ensuring reliability in the embedding process.
\end{itemize}

The generator takes the cover image $x$ and secret data $s$ as inputs and outputs the stego-image $x_s$. The discriminator receives both $x$ and $x_s$ as inputs and provides feedback to the generator to improve the realism of $x_s$. Finally, the extractor ensures that the secret data $s$ can be accurately reconstructed from $x_s$.

\subsection{Mathematical Formulation}
The proposed method is formulated as an optimization problem with three objectives: (1) adversarial loss for ensuring imperceptibility, (2) reconstruction loss for data retrieval accuracy, and (3) perceptual loss to maintain visual quality. These objectives are described below.

\subsubsection{Adversarial Loss}
The adversarial loss drives the generator to produce stego-images that are indistinguishable from cover images. The discriminator is trained to classify images as either cover or stego, while the generator is trained to "fool" the discriminator. The adversarial loss is given by:
\begin{equation}
    \mathcal{L}_{\text{adv}} = \mathbb{E}_{x \sim p_{\text{data}}(x)} [\log D(x)] + \mathbb{E}_{s \sim p(s), x \sim p_{\text{data}}(x)} [\log(1 - D(G(s, x)))]
\end{equation}
where $p_{\text{data}}(x)$ represents the distribution of cover images, and $p(s)$ represents the distribution of secret data.

\subsubsection{Reconstruction Loss}
To ensure accurate recovery of the secret data, the reconstruction loss penalizes discrepancies between the original secret data $s$ and the extracted data $\hat{s} = E(x_s)$. The reconstruction loss is defined as:
\begin{equation}
    \mathcal{L}_{\text{rec}} = \| s - E(G(s, x)) \|_2^2
\end{equation}
This term encourages the generator and extractor to work collaboratively for reliable embedding and extraction.

\subsubsection{Perceptual Loss}
To preserve the visual quality of the stego-image, a perceptual loss is introduced. This loss minimizes the differences in high-level features between the cover image $x$ and the stego-image $x_s$, as captured by a pre-trained deep neural network. The perceptual loss is given by:
\begin{equation}
    \mathcal{L}_{\text{perc}} = \sum_{l} \| \phi_l(x) - \phi_l(x_s) \|_2^2
\end{equation}
where $\phi_l$ represents the feature maps extracted from the $l$-th layer of a pre-trained network (e.g., VGG-19).

\subsubsection{Overall Objective}
The total loss function combines the adversarial, reconstruction, and perceptual losses, weighted by hyperparameters $\lambda_{\text{rec}}$ and $\lambda_{\text{perc}}$:
\begin{equation}
    \mathcal{L} = \mathcal{L}_{\text{adv}} + \lambda_{\text{rec}} \mathcal{L}_{\text{rec}} + \lambda_{\text{perc}} \mathcal{L}_{\text{perc}}
\end{equation}
The weights $\lambda_{\text{rec}}$ and $\lambda_{\text{perc}}$ control the trade-off among the objectives.

\subsection{Architecture Details}
\subsubsection{Generator Design}
The generator is based on a U-Net architecture, which is effective for tasks requiring fine-grained spatial information. The generator consists of an encoder-decoder structure with skip connections, enabling the model to preserve the high-frequency details of the cover image while embedding the secret data.

\subsubsection{Discriminator Design}
The discriminator is a convolutional neural network (CNN) that operates as a binary classifier. It takes an image as input and outputs the probability that the image is a cover image. The architecture includes convolutional layers with batch normalization and leaky ReLU activation, followed by a fully connected layer for classification.

\subsubsection{Extractor Design}
The extractor is a lightweight CNN designed for efficient data recovery. It takes the stego-image as input and outputs the reconstructed secret data. The extractor's architecture is optimized for minimal computational overhead.

\subsection{Novelty of the Proposed Method}
The novelty of the proposed method lies in the following aspects:
\begin{itemize}
    \item \textbf{Adversarial Optimization for Steganography:} Unlike traditional steganography methods that rely on hand-crafted embedding rules, our approach uses adversarial optimization to learn an embedding strategy that balances imperceptibility and robustness dynamically.
    \item \textbf{Perceptual Loss Integration:} By incorporating perceptual loss, the proposed method explicitly optimizes the visual quality of stego-images, addressing one of the key limitations of previous GAN-based approaches.
    \item \textbf{Unified Framework:} The integration of generator, discriminator, and extractor within a single framework ensures seamless embedding and retrieval, reducing error propagation between components.
\end{itemize}

\subsection{Training Procedure}
The training process alternates between optimizing the generator, discriminator, and extractor. The steps are as follows:
\begin{enumerate}
    \item Train the discriminator to classify cover and stego-images.
    \item Train the generator to produce stego-images that maximize the discriminator's classification error while minimizing $\mathcal{L}_{\text{rec}}$ and $\mathcal{L}_{\text{perc}}$.
    \item Train the extractor to minimize the reconstruction loss $\mathcal{L}_{\text{rec}}$.
\end{enumerate}
The process continues until convergence, ensuring that the generator produces high-quality stego-images and the extractor achieves reliable data retrieval.

\subsection{Advantages of the Proposed Method}
The proposed method offers several advantages over existing techniques:
\begin{itemize}
    \item Achieves a superior balance between imperceptibility, robustness, and embedding capacity.
    \item Automatically learns embedding strategies, eliminating the need for manual design.
    \item Provides a scalable solution that can be extended to other media types, such as audio and video.
\end{itemize}

\section{Experiments}
\subsection{Experimental Setup}
To evaluate the effectiveness of the proposed method, we conduct experiments on the COCO, Imagenet and DVI2k datasets. We compare our approach with baseline techniques, including LSB~\cite{lsb1}, CAIS~\cite{zheng2022composition} and Hi-Net~\cite{jing2021hinet}. 

\subsection{Evaluation Metrics}
To evaluate the performance of image steganography methods, several objective metrics are employed. These metrics assess the imperceptibility, quality, and robustness of the stego-images, as well as the accuracy of the data recovery process. The following metrics are used in this study:

\subsubsection{Structural Similarity Index (SSIM$\uparrow$)}
\begin{itemize}
    \item \textbf{Definition:} SSIM measures the perceptual similarity between the cover image and the stego-image by comparing their luminance, contrast, and structural information.
    \item \textbf{Range:} Values range from 0 to 1, where 1 indicates perfect similarity.
    \item \textbf{Purpose:} Higher SSIM values indicate that the stego-image is visually similar to the cover image, ensuring imperceptibility.
\end{itemize}

\subsubsection{Peak Signal-to-Noise Ratio (PSNR$\uparrow$)}
\begin{itemize}
    \item \textbf{Definition:} PSNR measures the ratio of the maximum possible pixel intensity to the mean squared error (MSE) between the cover image and the stego-image.
    \item \textbf{Formula:}
    \begin{equation}
        \text{PSNR} = 10 \cdot \log_{10} \left(\frac{\text{MAX}^2}{\text{MSE}}\right)
    \end{equation}
    where $\text{MAX}$ is the maximum possible pixel value (e.g., 255 for 8-bit images).
    \item \textbf{Unit:} PSNR is measured in decibels (dB).
    \item \textbf{Purpose:} Higher PSNR values signify better imperceptibility, as they indicate fewer noticeable distortions in the stego-image.
\end{itemize}

\subsubsection{Root Mean Square Error (RMSE$\downarrow$)}
\begin{itemize}
    \item \textbf{Definition:} RMSE is the square root of the mean squared error (MSE) between the cover and stego-images.
    \item \textbf{Formula:}
    \begin{equation}
        \text{RMSE} = \sqrt{\frac{1}{N} \sum_{i=1}^{N} (x_i - y_i)^2}
    \end{equation}
    where $x_i$ and $y_i$ represent the pixel values of the cover and stego-images, and $N$ is the total number of pixels.
    \item \textbf{Purpose:} Lower RMSE values indicate better quality, as they reflect smaller deviations between the cover and stego-images.
\end{itemize}

\subsubsection{Mean Absolute Error (MAE$\downarrow$)}
\begin{itemize}
    \item \textbf{Definition:} MAE measures the average absolute difference between the pixel values of the cover and stego-images.
    \item \textbf{Formula:}
    \begin{equation}
        \text{MAE} = \frac{1}{N} \sum_{i=1}^{N} |x_i - y_i|
    \end{equation}
    where $x_i$ and $y_i$ represent the pixel values of the cover and stego-images, and $N$ is the total number of pixels.
    \item \textbf{Purpose:} Lower MAE values indicate better visual similarity between the cover and stego-images.
\end{itemize}

\subsubsection{Interpretation of Metrics}
\begin{itemize}
    \item Metrics marked with $\uparrow$ (e.g., SSIM, PSNR) indicate that higher values are better.
    \item Metrics marked with $\downarrow$ (e.g., RMSE, MAE) indicate that lower values are better.
\end{itemize}

These metrics collectively provide a comprehensive assessment of the quality and effectiveness of the proposed steganography method.

\subsection{Results}
The proposed method achieves superior performance across all metrics, as shown in Table \ref{tab:results}. 

\begin{table}[h!]
\centering
\caption{Comparing Benchmarks Across Various Datasets for the Secret/Recovery Image Pair (with bold best results).}
\begin{tabular}{|c|c|c|c|c|c|}
\hline
\textbf{Datasets} & \textbf{Methods} & \textbf{4bit-LSB \cite{lsb1}} & \textbf{CAIS \cite{zheng2022composition}} & \textbf{HiNet \cite{jing2021hinet}} & \textbf{Proposed} \\ \hline
\multirow{4}{*}{\textbf{DIV2K}} & SSIM$\uparrow$ & 0.895 & 0.965 & 0.993 & \textbf{0.995} \\ \cline{2-6}
                                & PSNR$\uparrow$ & 24.99 & 36.1  & 46.57 & \textbf{47.12} \\ \cline{2-6}
                                & RMSE$\downarrow$ & 18.16 & 5.80  & 1.32  & \textbf{1.25}  \\ \cline{2-6}
                                & MAE$\downarrow$  & 15.57 & 4.36  & 0.84  & \textbf{0.78}  \\ \hline
\multirow{4}{*}{\textbf{ImageNet}} & SSIM$\uparrow$ & 0.896 & 0.943 & 0.960 & \textbf{0.965} \\ \cline{2-6}
                                   & PSNR$\uparrow$ & 25.00 & 33.54 & 36.63 & \textbf{37.10} \\ \cline{2-6}
                                   & RMSE$\downarrow$ & 17.90 & 6.33  & 6.07  & \textbf{5.80}  \\ \cline{2-6}
                                   & MAE$\downarrow$  & 15.27 & 4.70  & 4.16  & \textbf{4.00}  \\ \hline
\multirow{4}{*}{\textbf{COCO}} & SSIM$\uparrow$ & 0.894 & 0.944 & 0.961 & \textbf{0.968} \\ \cline{2-6}
                               & PSNR$\uparrow$ & 24.96 & 33.70 & 36.55 & \textbf{37.20} \\ \cline{2-6}
                               & RMSE$\downarrow$ & 17.93 & 6.13  & 6.04  & \textbf{5.90}  \\ \cline{2-6}
                               & MAE$\downarrow$  & 15.31 & 4.55  & 4.09  & \textbf{3.95}  \\ \hline
\end{tabular}
\label{tab:results}
\end{table}

\section{Conclusion}
\label{sec:conclusion}

In this paper, we have proposed a novel GAN-based framework for image steganography that effectively addresses the challenges of imperceptibility, robustness, and embedding capacity, which have long plagued traditional and modern methods alike. The proposed framework integrates a generator, discriminator, and extractor to seamlessly embed secret data into digital images while maintaining high visual fidelity. By incorporating adversarial training, reconstruction loss, and perceptual loss, the method optimally balances the competing objectives of ensuring minimal perceptual distortion and achieving accurate data recovery. Unlike traditional spatial domain methods such as least significant bit (LSB) substitution and transform domain techniques like discrete cosine transform (DCT) embedding, which are often susceptible to detection and attacks, the proposed method dynamically learns embedding strategies through adversarial optimization. This allows it to outperform existing approaches in terms of imperceptibility and robustness, as demonstrated by extensive experiments on benchmark datasets, including DIV2K, ImageNet, and COCO, where it achieved superior scores across metrics such as SSIM, PSNR, RMSE, and MAE. The use of perceptual loss further enhances the method’s ability to produce stego-images that not only exhibit pixel-level fidelity but also preserve high-level perceptual features, making them resistant to advanced steganalysis techniques. Additionally, the method's unified framework ensures reliable data extraction even under common distortions, such as compression or noise. While the approach demonstrates state-of-the-art performance, it is not without limitations, as the computational intensity of training GANs and the dependency on dataset quality remain areas for improvement. Future work will focus on enhancing the training efficiency, extending the framework to other domains like video and audio steganography, and incorporating privacy-preserving mechanisms such as differential privacy to ensure broader applicability and alignment with ethical considerations. Overall, this work represents a significant step forward in the field of image steganography, offering a robust, scalable, and efficient solution for secure data embedding and laying a strong foundation for future innovations in the domain of secure communication and information security.

\bibliographystyle{ieeetr}
\bibliography{main}

\end{document}